\documentclass[pre,twocolumn,longbibliography,superscriptaddress]{revtex4-1}
\usepackage{amsfonts}
\usepackage{amsmath}
\usepackage{amssymb}
\usepackage{graphicx}
\usepackage{color}
\usepackage{hyperref}

\begin{document}
\title{Comparative study of the discrete velocity  and lattice Boltzmann methods for rarefied gas flows through irregular channels}

\author{Scott Lindsay}
\address{James Weir Fluids Laboratory, Department of Mechanical and Aerospace Engineering, University of Strathclyde, Glasgow G1 1XJ, UK}
\author{Wei Su}
\address{Istituto di Nanotecnologia, Consiglio Nazionale delle Ricerche, 70126, Bari, Italy}
\author{Haihu Liu}
\address{School of Energy and Power Engineering, Xi'an Jiaotong University, 28 West Xianning Road, Xi'an 710049, China}
\author{Lei Wu}
\email{Correspoding author: lei.wu.100@strath.ac.uk}
\address{James Weir Fluids Laboratory, Department of Mechanical and Aerospace Engineering, University of Strathclyde, Glasgow G1 1XJ, UK}

\begin{abstract}


Rooted from the gas kinetics, the lattice Boltzmann method is a powerful tool in modeling hydrodynamics. In the past decade, it has been extended to simulate the rarefied gas flow beyond the Navier-Stokes level, either by using the high-order Gauss-Hermite quadrature, or by introducing the relaxation time that is a function of the gas-wall distance. While the former method, with a limited number of discrete velocities (i.e. D2Q36), is accurate up to the early transition flow regime, the latter method, with the same discrete velocities as that used in simulating hydrodynamics (i.e. D2Q9), is accurate up to the free-molecular flow regime in the Poiseuille flow between two parallel plates. This is quite astonishing in the sense that more discrete velocities are less accurate. In this paper, by solving the Bhatnagar-Gross-Krook kinetic equation accurately via the discrete velocity method, we find that the accuracy of the lattice Boltzmann method is reduced significantly in the simulation of rarefied gas flows through the rough surface and porous media. Our simulation results could serve as benchmarking cases for future development of the lattice Boltzmann method for  modeling and simulation of rarefied gas flows in complex geometries.

\end{abstract}
\maketitle

\date{text}

\section{Introduction}

When the mean free path $\lambda$ of gas molecules is comparable to or even larger than the characteristic flow length $H$, the Navier-Stokes (NS) equations derived from the continuum-fluid hypothesis fails, and the gas kinetic theory based on the fundamental Boltzmann equation for the velocity distribution function (VDF) of gaseous molecules
is required to describe rarefied gas dynamics. According to the Chapman-Enskog expansion, NS equations are only the first-order approximation in the Knudsen number ($\text{Kn}=\lambda/H$) to the Boltzmann equation~\cite{CE}. Therefore, they are valid in the continuum flow regime where $\text{Kn}\lesssim0.001$~\cite{Gad-el-Hak1999}.
As the Knudsen number increases,
higher-order (non-equilibrium) terms begin to dominate, and NS equations gradually loss validity; for instance, when $Kn$ is large, non-equilibrium terms embodied in the macroscopic mass, momentum and energy transport cannot be expressed in terms of the lower-order macroscopic quantities. The non-equilibrium effects not only cause noticeable velocity slip and temperature jump at solid surfaces in the slip flow regime ($0.001\lesssim\text{Kn}\lesssim0.1$), but also modify the constitutive relations in the transition ($0.1\lesssim\text{Kn}\lesssim10$) and free-molecular ($10\lesssim\text{Kn}$) flow regimes such that the Newton's law for stress and strain and Fourier's law for heat flux and temperature gradient do not hold anymore. This leads to a number of counterintuitive phenomena, including the thermal transpiration where gas molecules along solid surface move from the cold region to hot~\cite{Reynolds1879}, the Knudsen paradox where the dimensionless mass flow rate in Poiseuille flow could increase when the gas pressure decreases~\cite{Steckelmacher1999}, the temperature bimodality in the force-driven Poiseuille flow~\cite{Garcia1997TB}, the inverted velocity in cylindrical Couette flow~\cite{Aoki2003}, and the gas anti-resonance where the shear stress in an oscillating lid-driven cavity flow could be smaller than that of the one-dimensional (1D) Couette flow~\cite{WUjfm2014}.

In the past decades, due to the rapid development of microelectromechanical systems~\cite{Karniadakis2005a} and the shale gas revolution in North America~\cite{Wang2014Shale}, extensive works have been devoted to construct efficient numerical schemes to simulate gas flows at the micro scale, where the flow velocity is usually far smaller than the most probable speed of the gas molecules. In most of these applications, gas flows vary from the slip to the free-molecular flow regimes and the gas-surface interaction dominates flow behavior. High-fidelity numerical methods to solve the Boltzmann equation include the numerical kernel method~\cite{Ohwada_sone_1989}, the conservative projection-interpolation method~\cite{Tcheremissine2005}, the low-variance direct simulation Monte Carlo method~\cite{Homolle2007}, and the fast spectral method~\cite{lei_Jfm}, to name just a few. Due to the high computational cost, however, the Boltzmann equation is usually simplified by the Bhatnagar-Gross-Krook (BGK) equation under the single relaxation time (SRT) approximation~\cite{Bhatnagar1954}, which is often solved by the discrete velocity method (DVM) where the continuous molecular velocity space is represented by a few number of discrete velocities~\cite{ChuDVM1965,YangJCP1995,Sharipov1996PF,XuUGKS2010,GuoDUGKS2013}. Generally speaking, rarefied gas flows with higher values of $\text{Kn}$ need a larger number of discrete velocities to resolve the large variations/discontinuities in the VDF, for instance see the numerical examples in Refs.~\cite{lei_Jfm,AokiJFM2012}.

Rooted from the gas kinetics, the lattice Boltzmann method (LBM) is a popular and powerful tool in modeling the NS hydrodynamics and beyond. Historically, the SRT-LBM is firstly developed as an alternative solver for the NS equations. Since it uses a very limited but highly optimized number of discrete velocities, e.g. the D2Q9 scheme for 2D problems, the SRT-LBM can be viewed as a special type of DVM to solve the BGK equation~\cite{Xiaoyi1997}. To capture the non-equilibrium effects beyond the NS hydrodynamics, a rigorous procedure for obtaining high-order approximations to the BGK equation is proposed~\cite{ShanJFM2006} and tested in various canonical problems~\cite{KimJcp2008,MengJcp2011, MengPre2011,ShiPreR2011,SofoneaPre2012,SofoneaPre2014,ShiPre2015}. By expanding the VDF into high-order Gauss-Hermite polynomials or Gauss quadrature in the spherical coordinate system, it is found that, with a not very large number of discrete lattice velocities, e.g. D2Q36 or D2Q64, high-order LBM schemes accurately describe rarefied gas flows up to the early transition flow regime, i.e. $Kn\lesssim0.5$.


In addition to high-order quadrature, Zhang~\textit{et al.}~\cite{ZhangPre2006} attempted to capture the Knudsen layer structure in  rarefied gas flows within the framework of the SRT-LBM, but by introducing the gas kinetic boundary condition and the effective relaxation time that is a function of the gas-wall distance. In the pressure-driven flow, this ``wall-scaling'' approach provides a significant improvement for Knudsen numbers up to 0.5. Later, this scheme has also been applied to study the rarefied thermal~\cite{ZhangEpl2007} and oscillatory Couette flow~\cite{Tang2008} in the early transition flow regime with great success.

%

Further improvement has been achieved by Guo, Zheng and Shi~\cite{Guo2008Slip},  who developed a numerical scheme within the framework of multiple relaxation time (MRT) LBM. In addition to the ``wall-scaling'' of the relaxation time, the combined bounce-back and specular-reflection boundary condition is designed, such that the new scheme is equivalent to solve the NS equations with the second-order slip boundary condition in the slip flow regime. In the Poiseuille flow between two parallel plates, it is found that this MRT-LBM scheme is even able to predict the velocity profile and mass flow rate with good accuracy, in the whole transition flow regime. For this reason, the MRT-LBM has attracted significant attentions~\cite{Guo2008Slip,QliLBM2011,Rahman2016,ChenRough2016,YaoJAP2016}.

In the simulation of rarefied gas flows, it is quite astonishing that the MRT-LBM of Guo \textit{et al.} with fewer number of discrete velocities~\cite{Guo2008Slip} are more accurate than the LBM of higher-order quadrature~\cite{ShanJFM2006,KimJcp2008,MengJcp2011, MengPre2011,ShiPreR2011,SofoneaPre2012,SofoneaPre2014,ShiPre2015}. Although Guo~\textit{et al.} honestly and explicitly pointed out that their scheme is only designed and analyzed for plane walls~\cite{Guo2008Slip} and is hard to be extended to generalized gas-surface boundaries~\cite{GuoPRE2014}, it has been applied to study the rarefied gas flows through the microchannel with rough surface and complex porous media~\cite{ChenRough2016,YaoJAP2016}. However, whether this scheme works on walls with curvatures and other complex geometries or not is not clear.

In this paper, aiming to address the accuracy of the MRT-LBM in simulating rarefied gas flows in complex geometries, the BGK equation will be solved by the DVM, and numerical results will be compared to those of MRT-LBM for the Couette flow through rough microchannels~\cite{ChenRough2016} in Sec.~\ref{S_Couette}, and for the Poiseuille flow through porous media~\cite{YaoJAP2016} in  Sec.~\ref{S_Poiseuille}.


\section{Couette flow in rough microchannels}\label{S_Couette}

\begin{figure}[t]
 	\centering
 	\includegraphics[scale=0.45,viewport=30 290 615 545,clip=true]{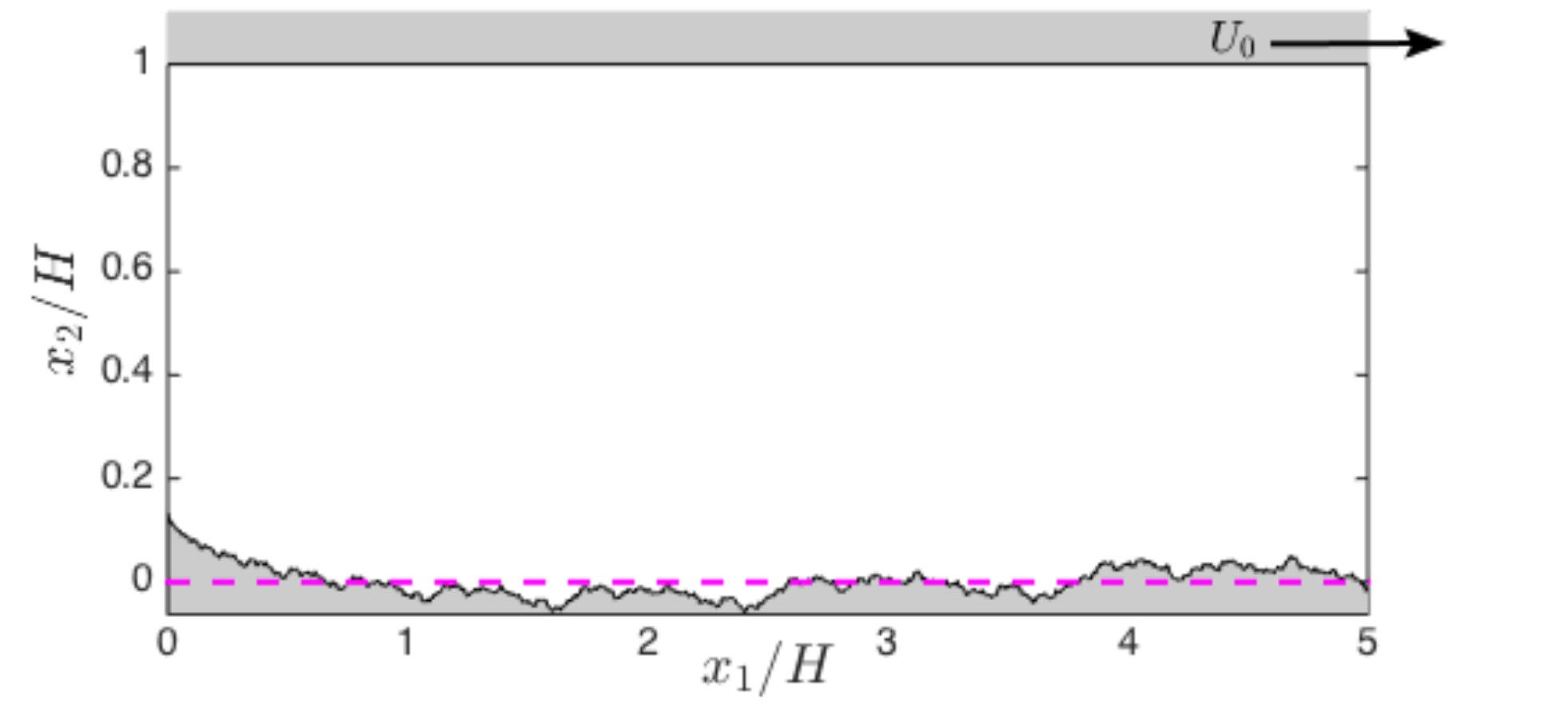}
 	\caption{Schematic of the Couette flow in a rough channel. Shaded regions are solid surfaces, while the gas flows are in the white region. }
 	\label{PRE_roughChannel}
\end{figure}

As Ref.~\cite{ChenRough2016}, we first consider the Couette flow between two plates with a distance of $H$, see Fig.~\ref{PRE_roughChannel}. The smooth top plate move in the $x_1$ direction with a speed $U_0$, while the stationary bottom plate has a roughness described by the following Weierstrass-Mandelbrot fractal function:
\begin{equation}\label{WM}
r(x_1/H)=A\sum_{n=n_1}^{\infty}\frac{\cos\left(2\pi\gamma^n{x_1/H}\right)}{\gamma^{(2-D)n}},
\end{equation}
where $D$ is the self-affine fractal dimension and $\gamma$ determines the frequency spectrum of the surface roughness. The aspect ratio of the microchannel is 5, and we choose $\gamma=1.5$ and $n_1=-4$ in Eq.~\eqref{WM}. The scaling parameter $A$ is used to adjust the surface roughness $\epsilon$:
\begin{equation}
\epsilon=\frac{\sigma}{H},
\end{equation}
where $\sigma$ is the root mean square of $r(x_1)$.

We are interested in the horizontal velocity profile of the gas and the mass flow rate at different level of gas rarefaction, when the steady state is reached.

\subsection{The gas kinetic theory and the DVM}

The fundamental Boltzmann equation describes the evolution of the VDF in dependence of spatial position and time. Macroscopic quantities are obtained from the velocity moments of the VDF. In this paper, we adopt the BGK equation~\cite{Bhatnagar1954} instead of the full Boltzmann collision operator. Numerical simulation for Poiseuille flow shows that the BGK equation can produce accurate mass flow rates~\cite{Sharipov2009,Doi2010}.

The BGK equation for the 2D problem may be written in the following form
\begin{equation}\label{Boltzmann}
\frac{\partial f}{\partial t}+\textbf{v}\cdot\frac{\partial f}{\partial \textbf{x}}=\frac{\sqrt{\pi}}{2\text{Kn}}(F_{eq}-f),
\end{equation}
where $f$ is the reduced two-velocity distribution function normalized by $\bar{p}/v_m^2mRT_0$ with $\bar{p}$ being the mean gas pressure at the reference temperature $T_0$ and $m$ the molecular mass, $\textbf{v}(v_1,v_2)$ is the molecular velocity normalized by the most probable speed $v_m=\sqrt{2RT_0}$ with $R$ being the gas constant, $\textbf{x}(x_1,x_2)$ is the spatial coordinate normalized by the height $H$, $t$ is the time normalized by $H/v_m$,  and $F_{eq}$ is the equilibrium VDF defined as
\begin{equation}
F_{eq}=\frac{\rho}{\pi{T}}{\exp(-\frac{|\textbf{v}-\textbf{u}|^2}{T})},
\end{equation}
where $\rho$ is the number density of gas molecules normalized by $\bar{p}/mRT_0$, $\textbf{u}(u_1,u_2)$ is the flow velocity normalized by $v_m$, and $T$ is the temperature normalized by $T_0$. Note that the Knudsen number is defined as $\text{Kn}=\lambda/H$, where the mean free path of gas molecules is related to its shear viscosity $\mu$ as
\begin{equation} \lambda=\frac{\mu(T_0)}{\bar{p}}\sqrt{\frac{\pi{RT_0}}{2}}.
\end{equation}

When the flow velocity is very small compared to $v_m$, we can linearize the VDF about the equilibrium state $f_{eq}$ as follows:
\begin{equation}
f=f_{eq}(1+h), \quad f_{eq}=\frac{\exp(-|\textbf{v}|^2)}{\pi},
\end{equation}
and the perturbation VDF $h(t,\textbf{x},\textbf{v})$ is governed by the following linearized BGK equation~\cite{GraurVacuum2012}:
\begin{equation}\label{BGK}
\begin{aligned}[b]
\frac{\partial h}{\partial t}+\textbf{v}\cdot\frac{\partial h}{\partial \textbf{x}}&=\mathcal{L}(\varrho,\textbf{u},\tau)-\frac{\sqrt{\pi}}{2\text{Kn}}h,\\
\mathcal{L}(\varrho,\textbf{u},\tau)&=\frac{\sqrt{\pi}}{2\text{Kn}}\left[\varrho+2\textbf{u}\cdot{\textbf{v}}+\tau\left(|\textbf{v}|^2-1\right)\right],
\end{aligned}
\end{equation}
where
$\varrho=\int{h}f_{eq}\mathrm{d}\textbf{v}$ is the perturbed number density and $\tau=\int |\textbf{v}|^2{h}f_{eq}\mathrm{d}\textbf{v}-\varrho$ is the perturbed temperature. The velocity $\textbf{u}$ is also calculated from $h$, i.e. $u_1=\int v_1{h}f_{eq}\mathrm{d}\textbf{v}$ and $u_2=\int v_2{h}f_{eq}\mathrm{d}\textbf{v}$ respectively.

The kinetic equation~\eqref{BGK} has to be solved together with the gas kinetic boundary condition which determines the VDF of the reflected gas molecules at the surface in terms of that of the incident molecules. At the inlet and outlet of the computational domain, the periodic boundary condition is used, while at the solid surface, the diffuse boundary condition is used~\cite{Maxwell1879}. That is, at the moving smooth plate located at $x_2=1$, we have
\begin{equation}\label{diffuse1}
\begin{aligned}[b]
h(x_1,1,\textbf{v})={2}{\sqrt\pi} \int_{v_2'>0} v_2'h(x_1,1,\textbf{v}')f_{eq}(\textbf{v}') \mathrm{d}\textbf{v}'+\frac{2U_0}{v_m}v_1, 
\end{aligned}
\end{equation}
while at the bottom rough plate located,
\begin{equation}\label{diffuse2}
h(x_1,x_2,\textbf{v})={2}{\sqrt\pi} \int_{v_n'<0} |v_n'| h(x_1,x_2,\textbf{v}')f_{eq}(\textbf{v}') \mathrm{d}\textbf{v}',
\end{equation}
where $v_n$ is the normal velocity vector at the solid surface pointing into the fluid regime.

Note that the Weierstrass-Mandelbrot fractal surface is not differentiable, therefore, the rough surface is approximated by the ``stair case" in the numerical simulation, as used in the MRT-LBM simulation~\cite{ChenRough2016}. Also, we take $U_0/v_m=1$, so the horizontal flow velocity is calculated as
\begin{equation}
u_1(x_1,x_2)=U_0\int{v_1f(x_1,x_2,\textbf{v})}\mathrm{d}\textbf{v},
\end{equation}
while the average mass flow rate, which is normalized by $\bar{p}HU_0/RT_0$, is given by
\begin{equation}
Q=\frac{\iiint{v_1f(x_1,x_2,\textbf{v})}\mathrm{d}\textbf{v}\mathrm{d}x_1\mathrm{d}x_2}{5}.
\end{equation}


In numerical simulations, the  molecular velocity space $\textbf{v}$ is represented by discrete velocities. To capture the discontinuities in the VDF at large $\text{Kn}$, $v_1$ and $v_2$ are represented by $N_v$ non-uniform points in each direction~\cite{lei_Jfm}:
\begin{equation}\label{nonuniform_v}
v_{1,2}=\frac{4}{(N_{v}-1)^3}(-N_{v}+1,-N_{v}+3,\cdots,{N_{v}-1})^3,
\end{equation}
where most of the discrete velocities are located near $v_{1,2}=0$. We choose $N_v=32$ in all our simulations.

The physical space is also discretized into Cartesian grids. In the $x_1$ direction, 1500 equidistant points are used. In the $x_2$ direction, the rough region $\min(r)\le{}x_2<\max(r)$ is discretized by 60 equidistant points, while the rest is discretized by 100 equidistant points.

As we are interested in the steady-state solution, the derivative with respect to the time is omitted in Eq.~\eqref{BGK}, and the resulting equation is solved by the following iterative method:
\begin{equation}\label{iteration}
\frac{\sqrt{\pi}}{2\text{Kn}}h^{(n+1)}+\textbf{v}\cdot\frac{\partial h^{(n+1)}}{\partial \textbf{x}}=\mathcal{L}(\varrho^{(n)},\textbf{u}^{(n)},\tau^{(n)}),
\end{equation}
where the superscript $(n)$ and $(n+1)$ represent two consecutive iteration steps, and the spatial derivatives $\partial h/\partial x$ and $\partial h/\partial y$ are approximated by the second-order upwind finite difference. The iteration is terminated when the relative error in the horizontal flow velocity between two consecutive iteration steps is less than $10^{-5}$.

\subsection{Numerical results}

 \begin{figure}
 	\centering
 	\includegraphics[scale=0.45,viewport=60 230 515 585,clip=true]{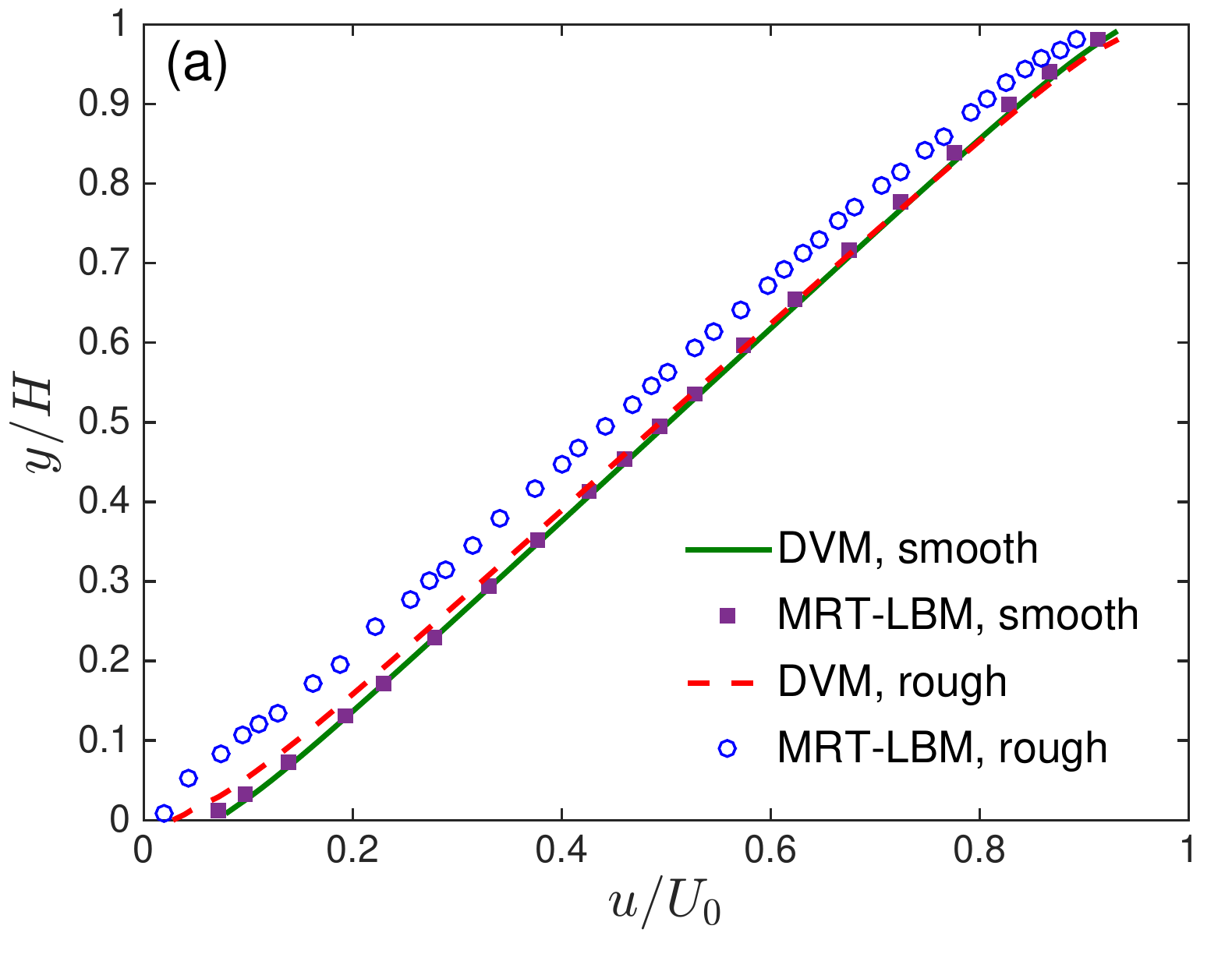}
 	\includegraphics[scale=0.45,viewport=60 240 515 585,clip=true]{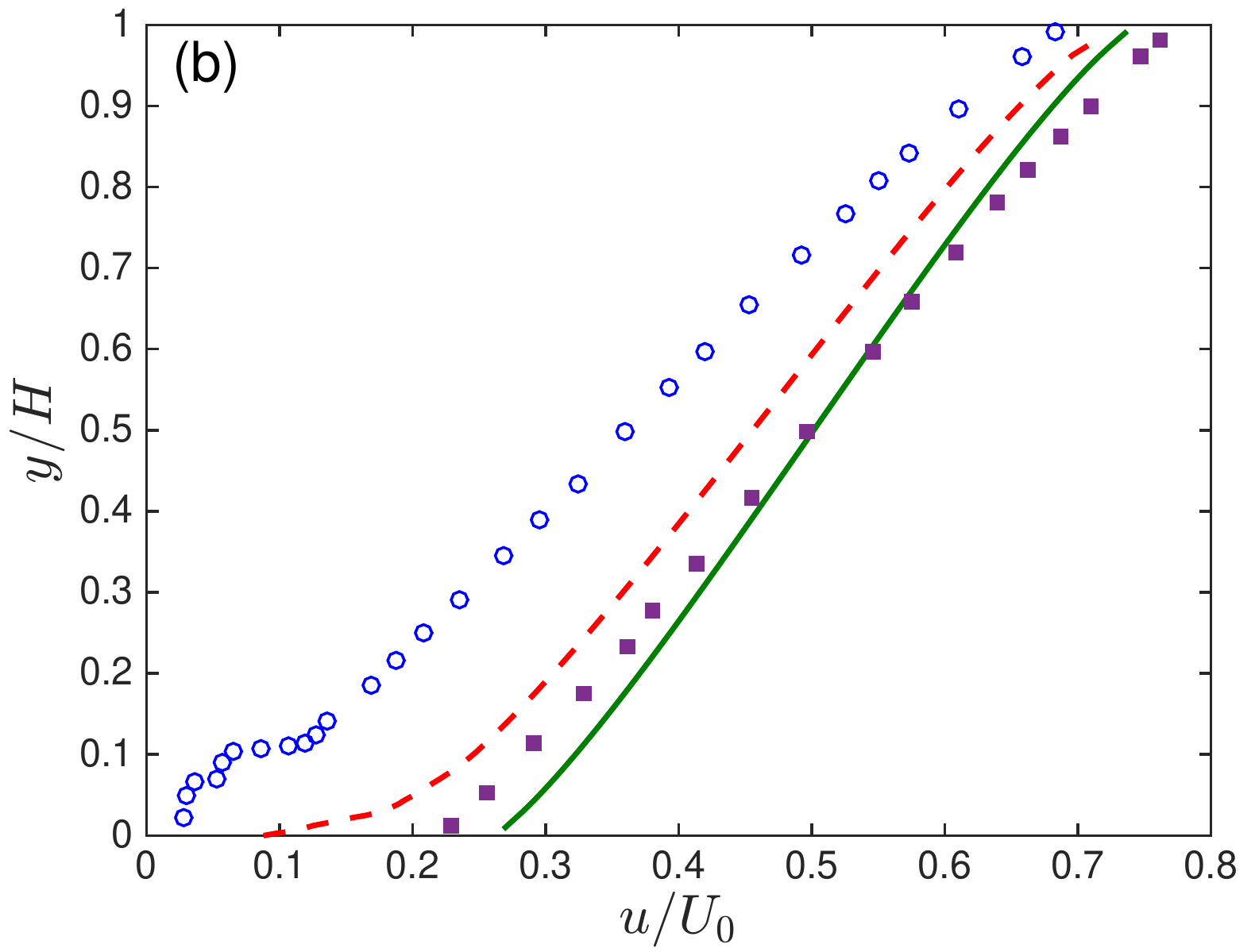}
 	\caption{Comparisons in the average streamwise velocity profile of the Couette flow in microchannels between the DVM and MRT-LBM~\cite{ChenRough2016}, when $D=1.5$, $\epsilon=2\%$, (a) $\text{Kn}=0.1$ and (b) $\text{Kn}=1$. ``Smooth'' here means that both plates are smooth, while ``rough'' means that only the bottom plate has a rough surface, see Fig.~\ref{PRE_roughChannel}. }
 	\label{PRE_vel}
 \end{figure}

We first compare the velocity profiles between the numerical results obtained from the DVM and MRT-LBM, when $\text{Kn}=0.1$ and 1. When both plates are smooth, from Fig.~\ref{PRE_vel} we find that MRT-LBM is accurate at $\text{Kn}=0.1$. However, at $\text{Kn=1}$, the slip velocity at the bottom surface from the MRT-LBM simulation is about 15~\% smaller than that from the DVM. Nevertheless, the mass flow rates from both numerical schemes are quite close, as the MRT-LBM predicts a larger velocity close to the top plate. When the bottom surface is rough, the accuracy of the MTR-LBM is reduced significantly. For example, at $\text{Kn}=1$ and $x_2=0.1$ (i.e. the rough region), the average horizontal flow velocity from the DVM simulation is two and a half times larger than that from the MRT-LBM. Possible reasons are, (i) the combined bounce-back and specular-reflection boundary condition is only derived in a large flat plate, which may not work in a rough surface, and (ii) for rough surface, the ``wall-scaling'' of the relaxation time may not work properly. Actually, it has been observed that, the extend NS equations incorporating a wall-scaling model cannot give a correct prediction in velocity profiles even for a simple circular cylinder flow when Kn$>0.4$~\cite{GuoPRE2014}.

\begin{figure}[t]
 	\centering
 	\includegraphics[scale=0.45,viewport=60 230 515 585,clip=true]{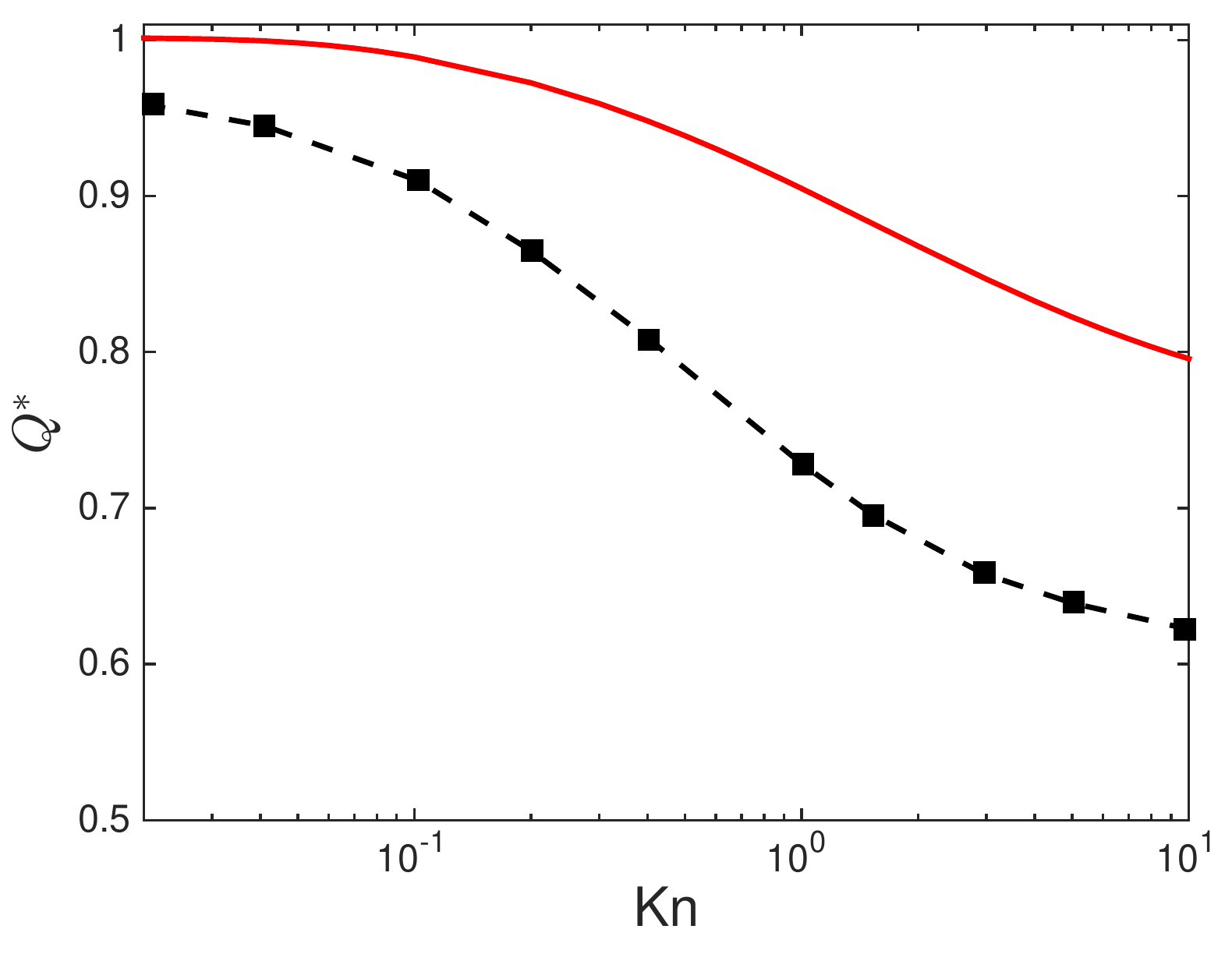}
 	\caption{The ratio between the mass flow rates in the rough and smooth channels when $D=1.5$ and $\epsilon=2\%$. Solid line: DVM. Squares: MRT-LBM from~\cite{ChenRough2016}. }
 	\label{PRE_massflow}
\end{figure}

We then consider the influence of the surface roughness on the  mass flow rate. As Ref.~\cite{ChenRough2016}, we study how the following ratio
\begin{equation}
 Q^\ast=\frac{Q_r}{Q_s},
\end{equation}
change with the Knudsen number. Here $Q_r$ and $Q_s$ are the average mass flow rates through the channels with rough and smooth surfaces, respectively. For the smooth channel with diffuse boundary condition, our numerical simulations show that $Q_s$ is always 0.5 in the whole flow regime. This can actually be proven in the continuum and slip flow regimes using the NS equations with slip boundary condition, and in the free-molecular flow regime using  the gas kinetic theory. The numerical results for fractal surface with $D=1.5$ and $\epsilon=2\%$ are shown in Fig.~\ref{PRE_massflow}. Clearly, the accuracy of MRT-LBM decreases as the Knudsen number increases. At $\text{Kn}=10$, $Q^\ast$ obtained from MRT-LBM is about 25\% smaller than that from the DVM. This error, however, is in fact very large by taking into account that the value of $Q$ only decreases by 20\% from the continuum flow regime to the free-molecular flow regime with $\text{Kn}=10$. In this sense, the MRT-LBM overestimates the decrease of $Q$ by twice.

We also checked the mass flow rate when the fractal dimension $D$ in Eq.~\eqref{WM} takes different values. Similar behaviors as that in Fig.~\ref{PRE_massflow} are observed (not shown). These examples show that the MRT-LBM is not reliable in simulating Couette flow in rough microchannels.

\section{Poiseuille flow in porous media}\label{S_Poiseuille}

\begin{figure}[t]
	\centering
	\includegraphics[scale=0.5,viewport=160 330 465 525,clip=true]{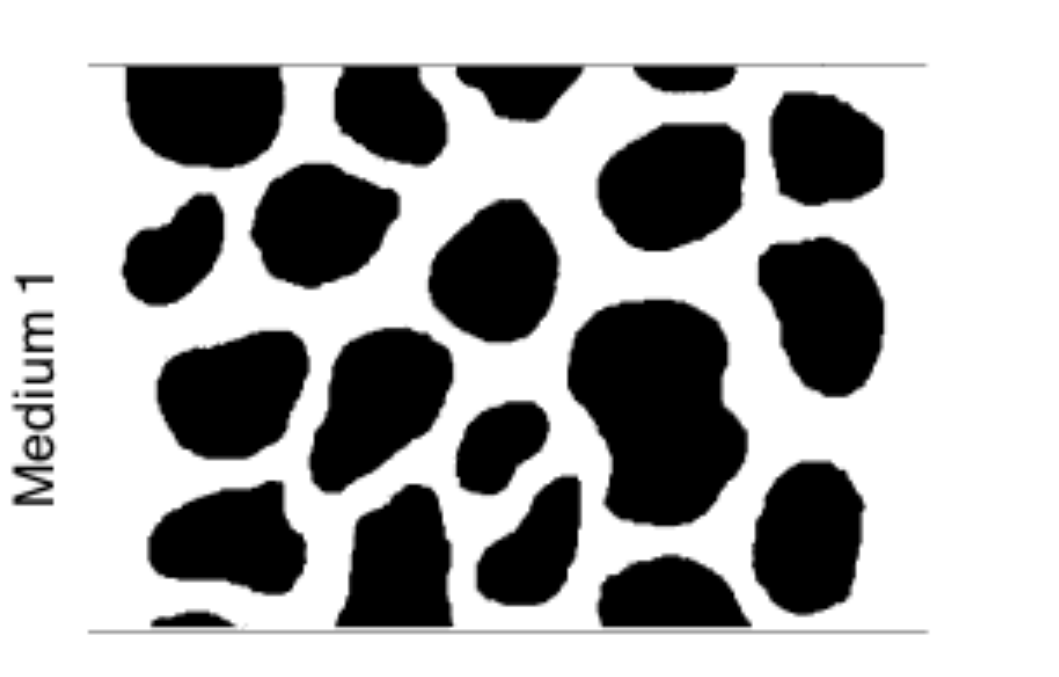}\\
	\includegraphics[scale=0.47,viewport=60 340 655 535,clip=true]{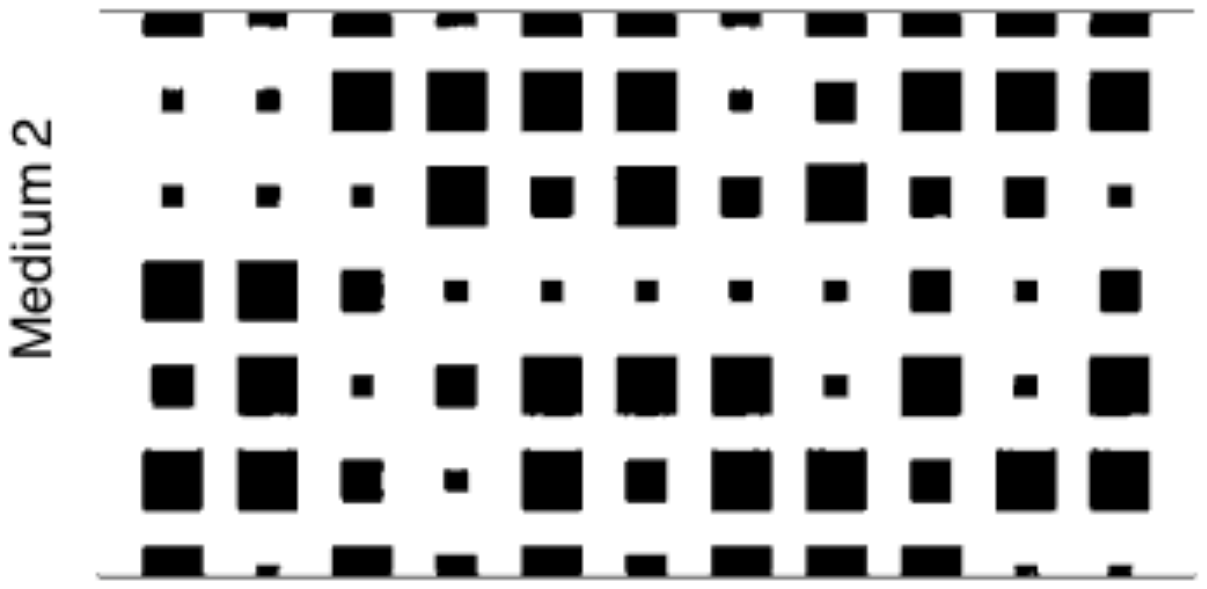}
	\\
	\includegraphics[scale=0.47,viewport=60 340 635 545,clip=true]{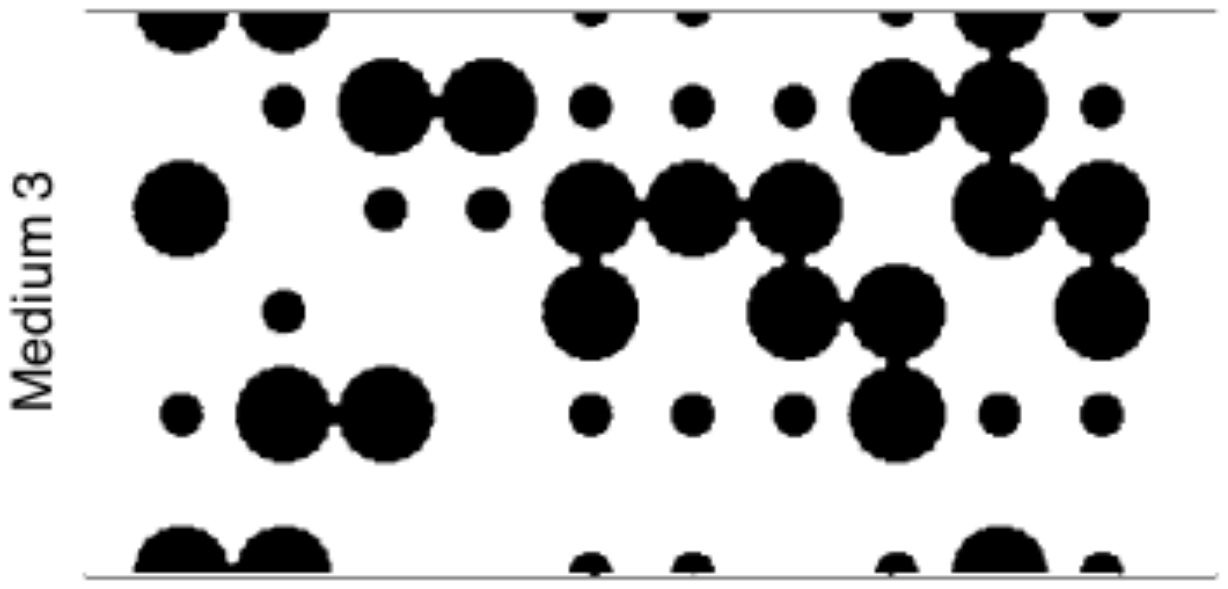}
	\caption{Different porous media used in the numerical  simulation~\cite{YaoJAP2016}. The gas flows from left to right. }
	\label{Medium_Pois}
\end{figure}

In this section, we further access the accuracy of the MRT-LBM in simulating rarefied Poiseuille flows through three porous medium (Medium 1 with irregular solids; Medium 2 with square solids; Medium 3 with circular solids), as shown in Fig.~\ref{Medium_Pois}. The vertical size of each porous medium is assumed to be $H$. This kind of research becomes popular due to the shale gas revolution in North America, and LBM is widely applied to find the apparent permeability (a parameter describing how fast the gas can be extracted) of porous media. We will find the variation of the apparent permeability with respect to the Knudsen number using the DVM, and access the accuracy of MRT-LBM. 

The DVM to solve the linearized BGK equation for this problem can be found in Ref.~\cite{LeiComment2017} where the periodic boundary condition is used. We have also tested that the pressure boundary condition adopted in Ref.~\cite{YaoJAP2016} leads to the same value of apparent permeability as long as the pressure difference is very small.

According to Ref.~\cite{LeiComment2017}, the apparent gas permeability $k_a$, which is normalized by $H^2$, is calculated by
\begin{equation}
k_a=\sqrt{\frac{4}{\pi}}Q\text{Kn},
\end{equation}
where $Q$ is the dimensionless mass flow rate
 \begin{equation}
Q=\frac{\iiint{v_1f(x_1,x_2,\textbf{v})}\mathrm{d}\textbf{v}\mathrm{d}x_1\mathrm{d}x_2}{L},
\end{equation}
with $L$ being length of the computational domain along the flow direction.

Since different porous medium has different distribution of pore radius, it will be useful to define the characteristic length as
\begin{equation}
H^\ast=H\sqrt{\frac{12k_\infty}{\phi}},
\end{equation}
where $\phi$ is the porosity of the porous medium and $k_\infty$ is the intrinsic permeability, i.e. the apparent permeability when $\text{Kn}\rightarrow0$. Hence the equivalent Knudsen number is defined as
\begin{equation}
	\text{Kn}^\ast=\frac{\lambda}{H^\ast}=\text{Kn}\frac{H}{H^\ast}=\text{Kn}\sqrt{\frac{\phi}{12k_\infty}}.
\end{equation}

\begin{figure}[t]
	\centering
	\includegraphics[scale=0.5,viewport=50 250 565 580,clip=true]{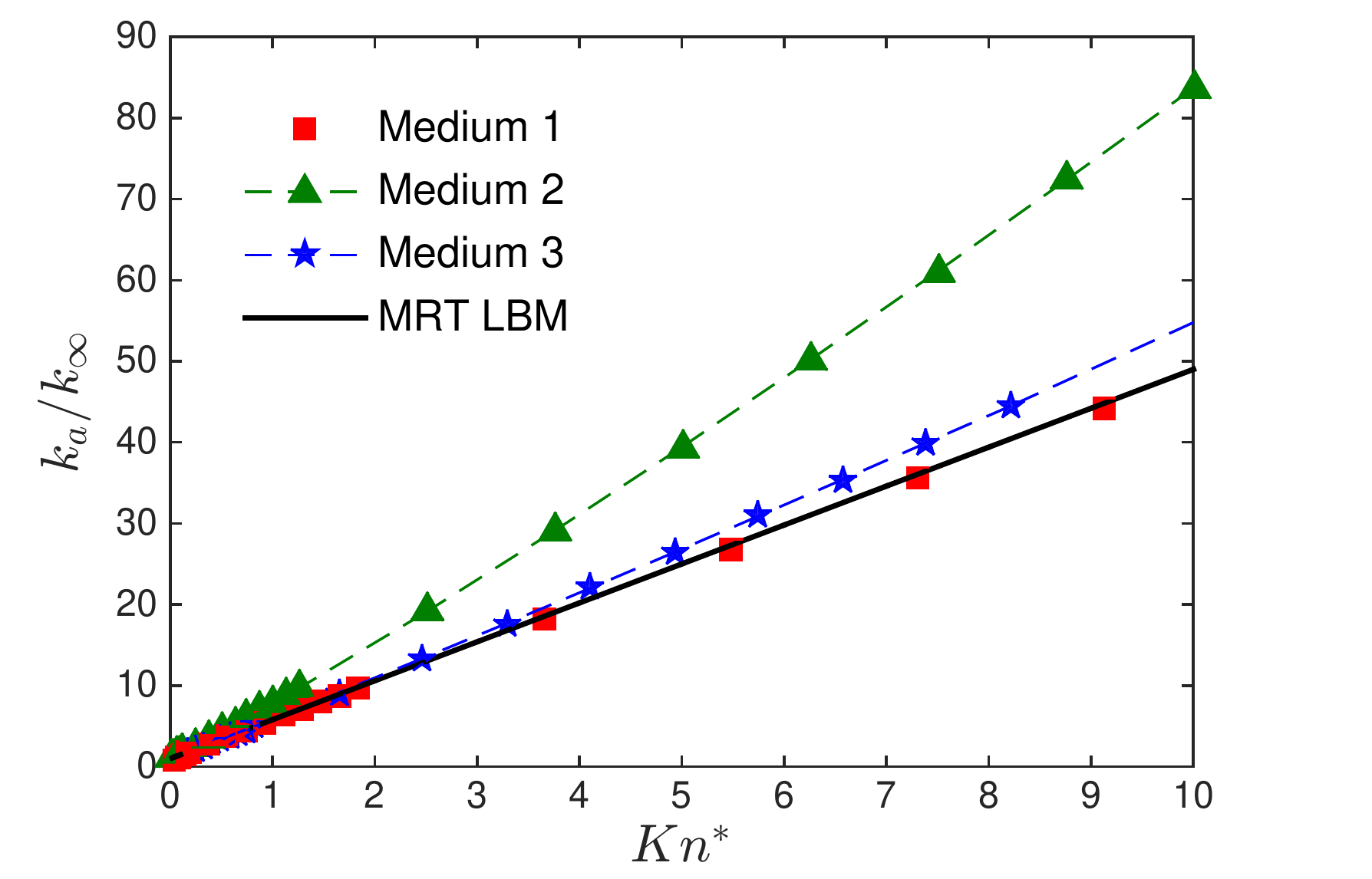}
	\caption{The ratio of the apparent gas permeability $k_a$ to the intrinsic permeability $k_\infty$ as a function of the equivalent Knudsen number. For porous media 1, 2, and 3, results from the MRT-LBM simulations~\cite{YaoJAP2016} fall into the black solid line when $\text{Kn}^\ast>1$. }
	\label{AGP}
\end{figure}

The comparison on the apparent gas permeability of the three porous media at different $Kn^\ast$ between the numerical results from the DVM and MRT-LBM is presented in Fig.~\ref{AGP}. Our DVM simulations show that, for the porous media 1 and 3, the ratio of the apparent gas permeability to the intrinsic permeability, i.e. $k_a/k_{\infty}$, is nearly proportional to the equivalent Knudsen number $\text{Kn}^\ast$, while for the porous medium 2, $k_a/k_{\infty}$ is a convex function of $Kn^\ast$. For the three porous media, surprisingly, the results from the MRT-LBM simulations fall into the same line when $\text{Kn}^\ast>1$. It is interesting to note that, for the porous medium 1 both DVM and MRT-LBM results agree with each other very well, while for the porous medium 3, the apparent permeability from DVM is only larger than those from MRT-LBM by about 10~\%. However, for the porous media 2, the apparent permeabilities predicted from DVM are nearly twice as  large as those from MRT-LBM.

Therefore, in general, the MTR-LBM cannot describe the variation of the apparent permeability with respect to the shape and configuration of pores in porous media, and should not viewed as an accurate numerical method to simulate rarefied gas flows in porous media.

\section{Conclusions}

In summary, we provided a systematic assessment of the accuracy of the MRT-LBM method in simulating rarefied gas flows in rough microchannels and complex porous media, over a wide range of the Knudsen number. By solving the linearized Bhatnagar-Gross-Krook kinetic equation accurately via the discrete velocity method, we have shown that:
\begin{enumerate}
	\item for Couette flow in microchannels, the MRT-LBM can predict the velocity profile with high accuracy in slip flow regime, when the two surfaces are  smooth. However, as the Knudsen number increases or surface roughness emerges, the MRT-LBM losses accuracy in predicting the velocity slip and mass flow rate: it even becomes very inaccurate when the Knudsen number is one.
	\item for Poiseuille flow through porous media, the MRT-LBM cannot describe the variation of the apparent permeability with respect to the shape and configuration of pores. In one case considered, it underestimates the apparent permeability by nearly twice.
\end{enumerate}

Therefore, for rarefied flows through rough surface and complex porous media, the MRT-LBM should not be viewed as an accurate method. Possible reasons for the failure of the MRT-LBM include (i) the combined bounce-back and specular-reflection boundary condition is only derived in a large flat plate, which may not work in a rough surface, and (ii) for rough surface, the ``wall-scaling'' of the relaxation time may not work properly.
Our simulation results from the discrete velocity method could serve as benchmarking cases for future development of the LBM for modeling and simulation of rarefied gas flows in complex geometry.

\vskip 0.5cm
L.W. and H.L. acknowledge the financial support of the joint project from the Royal Society of Edinburgh and National Natural Science Foundation of China.

\bibliography{Bib}

\begin{thebibliography}{43}%
\makeatletter
\providecommand \@ifxundefined [1]{%
 \@ifx{#1\undefined}
}%
\providecommand \@ifnum [1]{%
 \ifnum #1\expandafter \@firstoftwo
 \else \expandafter \@secondoftwo
 \fi
}%
\providecommand \@ifx [1]{%
 \ifx #1\expandafter \@firstoftwo
 \else \expandafter \@secondoftwo
 \fi
}%
\providecommand \natexlab [1]{#1}%
\providecommand \enquote  [1]{``#1''}%
\providecommand \bibnamefont  [1]{#1}%
\providecommand \bibfnamefont [1]{#1}%
\providecommand \citenamefont [1]{#1}%
\providecommand \href@noop [0]{\@secondoftwo}%
\providecommand \href [0]{\begingroup \@sanitize@url \@href}%
\providecommand \@href[1]{\@@startlink{#1}\@@href}%
\providecommand \@@href[1]{\endgroup#1\@@endlink}%
\providecommand \@sanitize@url [0]{\catcode `\\12\catcode `\$12\catcode
  `\&12\catcode `\#12\catcode `\^12\catcode `\_12\catcode `\%12\relax}%
\providecommand \@@startlink[1]{}%
\providecommand \@@endlink[0]{}%
\providecommand \url  [0]{\begingroup\@sanitize@url \@url }%
\providecommand \@url [1]{\endgroup\@href {#1}{\urlprefix }}%
\providecommand \urlprefix  [0]{URL }%
\providecommand \Eprint [0]{\href }%
\providecommand \doibase [0]{http://dx.doi.org/}%
\providecommand \selectlanguage [0]{\@gobble}%
\providecommand \bibinfo  [0]{\@secondoftwo}%
\providecommand \bibfield  [0]{\@secondoftwo}%
\providecommand \translation [1]{[#1]}%
\providecommand \BibitemOpen [0]{}%
\providecommand \bibitemStop [0]{}%
\providecommand \bibitemNoStop [0]{.\EOS\space}%
\providecommand \EOS [0]{\spacefactor3000\relax}%
\providecommand \BibitemShut  [1]{\csname bibitem#1\endcsname}%
\let\auto@bib@innerbib\@empty
\bibitem [{\citenamefont {Chapman}\ and\ \citenamefont {Cowling}(1970)}]{CE}%
  \BibitemOpen
  \bibfield  {author} {\bibinfo {author} {\bibfnamefont {S.}~\bibnamefont
  {Chapman}}\ and\ \bibinfo {author} {\bibfnamefont {T.G.}\ \bibnamefont
  {Cowling}},\ }\href@noop {} {\emph {\bibinfo {title} {{The Mathematical
  Theory of Non-uniform Gases}}}}\ (\bibinfo  {publisher} {Cambridge University
  Press},\ \bibinfo {year} {1970})\BibitemShut {NoStop}%
\bibitem [{\citenamefont {Gad{\--}el{\--}Hak}(1999)}]{Gad-el-Hak1999}%
  \BibitemOpen
  \bibfield  {author} {\bibinfo {author} {\bibfnamefont {M.}~\bibnamefont
  {Gad{\--}el{\--}Hak}},\ }\bibfield  {title} {\enquote {\bibinfo {title} {The
  fluid mechanics of microdevices - the {F}reeman {S}cholar {l}ecture},}\
  }\href@noop {} {\bibfield  {journal} {\bibinfo  {journal} {J. Fluids Eng.}\
  }\textbf {\bibinfo {volume} {121}},\ \bibinfo {pages} {5--33} (\bibinfo
  {year} {1999})}\BibitemShut {NoStop}%
\bibitem [{\citenamefont {Reynolds}(1879)}]{Reynolds1879}%
  \BibitemOpen
  \bibfield  {author} {\bibinfo {author} {\bibfnamefont {O.}~\bibnamefont
  {Reynolds}},\ }\bibfield  {title} {\enquote {\bibinfo {title} {On certain
  dimensional properties of matter in the gaseous state},}\ }\href@noop {}
  {\bibfield  {journal} {\bibinfo  {journal} {Phil. Trans. R. Soc. Lond.}\
  }\textbf {\bibinfo {volume} {170}},\ \bibinfo {pages} {727--845} (\bibinfo
  {year} {1879})}\BibitemShut {NoStop}%
\bibitem [{\citenamefont {Steckelmacher}(1999)}]{Steckelmacher1999}%
  \BibitemOpen
  \bibfield  {author} {\bibinfo {author} {\bibfnamefont {W.}~\bibnamefont
  {Steckelmacher}},\ }\bibfield  {title} {\enquote {\bibinfo {title} {Knudsen
  flow 75 years on: the current state of the art for flow of rarefied gases in
  tubes and systems},}\ }\href@noop {} {\bibfield  {journal} {\bibinfo
  {journal} {Rep. Prog. Phys.}\ }\textbf {\bibinfo {volume} {49}},\ \bibinfo
  {pages} {1083--1107} (\bibinfo {year} {1999})}\BibitemShut {NoStop}%
\bibitem [{\citenamefont {Mansour}\ \emph {et~al.}(1997)\citenamefont
  {Mansour}, \citenamefont {Baras},\ and\ \citenamefont
  {Garcia}}]{Garcia1997TB}%
  \BibitemOpen
  \bibfield  {author} {\bibinfo {author} {\bibfnamefont {M.~M.}\ \bibnamefont
  {Mansour}}, \bibinfo {author} {\bibfnamefont {F.}~\bibnamefont {Baras}}, \
  and\ \bibinfo {author} {\bibfnamefont {A.~L.}\ \bibnamefont {Garcia}},\
  }\bibfield  {title} {\enquote {\bibinfo {title} {On the validity of
  hydrodynamics in plane poiseuille flows},}\ }\href@noop {} {\bibfield
  {journal} {\bibinfo  {journal} {Physica A}\ }\textbf {\bibinfo {volume}
  {240}},\ \bibinfo {pages} {255} (\bibinfo {year} {1997})}\BibitemShut
  {NoStop}%
\bibitem [{\citenamefont {Aoki}\ \emph {et~al.}(2003)\citenamefont {Aoki},
  \citenamefont {Yoshida}, \citenamefont {Nakanishi},\ and\ \citenamefont
  {Garcia}}]{Aoki2003}%
  \BibitemOpen
  \bibfield  {author} {\bibinfo {author} {\bibfnamefont {K.}~\bibnamefont
  {Aoki}}, \bibinfo {author} {\bibfnamefont {H.}~\bibnamefont {Yoshida}},
  \bibinfo {author} {\bibfnamefont {T.}~\bibnamefont {Nakanishi}}, \ and\
  \bibinfo {author} {\bibfnamefont {A.~L.}\ \bibnamefont {Garcia}},\ }\bibfield
   {title} {\enquote {\bibinfo {title} {Inverted velocity profile in the
  cylindrical {C}ouette flow of a rarefied gas},}\ }\href@noop {} {\bibfield
  {journal} {\bibinfo  {journal} {Physical Review E}\ }\textbf {\bibinfo
  {volume} {68}},\ \bibinfo {pages} {016302} (\bibinfo {year}
  {2003})}\BibitemShut {NoStop}%
\bibitem [{\citenamefont {Wu}\ \emph {et~al.}(2014{\natexlab{a}})\citenamefont
  {Wu}, \citenamefont {Reese},\ and\ \citenamefont {Zhang}}]{WUjfm2014}%
  \BibitemOpen
  \bibfield  {author} {\bibinfo {author} {\bibfnamefont {L.}~\bibnamefont
  {Wu}}, \bibinfo {author} {\bibfnamefont {J.~M.}\ \bibnamefont {Reese}}, \
  and\ \bibinfo {author} {\bibfnamefont {Y.~H.}\ \bibnamefont {Zhang}},\
  }\bibfield  {title} {\enquote {\bibinfo {title} {Oscillatory rarefied gas
  flow inside rectangular cavities},}\ }\href@noop {} {\bibfield  {journal}
  {\bibinfo  {journal} {J. Fluid Mech.}\ }\textbf {\bibinfo {volume} {748}},\
  \bibinfo {pages} {350--367} (\bibinfo {year}
  {2014}{\natexlab{a}})}\BibitemShut {NoStop}%
\bibitem [{\citenamefont {Karniadakis}\ \emph {et~al.}(2005)\citenamefont
  {Karniadakis}, \citenamefont {Beskok},\ and\ \citenamefont
  {Aluru}}]{Karniadakis2005a}%
  \BibitemOpen
  \bibfield  {author} {\bibinfo {author} {\bibfnamefont {G.}~\bibnamefont
  {Karniadakis}}, \bibinfo {author} {\bibfnamefont {A.}~\bibnamefont {Beskok}},
  \ and\ \bibinfo {author} {\bibfnamefont {N.}~\bibnamefont {Aluru}},\
  }\href@noop {} {\emph {\bibinfo {title} {Microflows and Nanoflows:
  Fundamentals and Simulation}}}\ (\bibinfo  {publisher} {Springer},\ \bibinfo
  {year} {2005})\BibitemShut {NoStop}%
\bibitem [{\citenamefont {Wang}\ \emph {et~al.}(2014)\citenamefont {Wang},
  \citenamefont {Chen}, \citenamefont {Jha},\ and\ \citenamefont
  {Rogers}}]{Wang2014Shale}%
  \BibitemOpen
  \bibfield  {author} {\bibinfo {author} {\bibfnamefont {Q.}~\bibnamefont
  {Wang}}, \bibinfo {author} {\bibfnamefont {X.}~\bibnamefont {Chen}}, \bibinfo
  {author} {\bibfnamefont {A.}~\bibnamefont {Jha}}, \ and\ \bibinfo {author}
  {\bibfnamefont {H.}~\bibnamefont {Rogers}},\ }\bibfield  {title} {\enquote
  {\bibinfo {title} {{Natural gas from shale formation - The evolution,
  evidences and challenges of shale gas revolution in United States}},}\
  }\href@noop {} {\bibfield  {journal} {\bibinfo  {journal} {Renew. Sust.
  Energ. Rev.}\ }\textbf {\bibinfo {volume} {30}},\ \bibinfo {pages} {1--28}
  (\bibinfo {year} {2014})}\BibitemShut {NoStop}%
\bibitem [{\citenamefont {Ohwada}\ \emph {et~al.}(1989)\citenamefont {Ohwada},
  \citenamefont {Sone},\ and\ \citenamefont {Aoki}}]{Ohwada_sone_1989}%
  \BibitemOpen
  \bibfield  {author} {\bibinfo {author} {\bibfnamefont {T.}~\bibnamefont
  {Ohwada}}, \bibinfo {author} {\bibfnamefont {Y.}~\bibnamefont {Sone}}, \ and\
  \bibinfo {author} {\bibfnamefont {K}~\bibnamefont {Aoki}},\ }\bibfield
  {title} {\enquote {\bibinfo {title} {{Numerical analysis of the Poiseuille
  and thermal transpiration flows between two parallel plates on the basis of
  the Boltzmann equation for hard sphere molecules}},}\ }\href@noop {}
  {\bibfield  {journal} {\bibinfo  {journal} {Phys. Fluids}\ }\textbf {\bibinfo
  {volume} {1}},\ \bibinfo {pages} {2042} (\bibinfo {year} {1989})}\BibitemShut
  {NoStop}%
\bibitem [{\citenamefont {Tcheremissine}(2005)}]{Tcheremissine2005}%
  \BibitemOpen
  \bibfield  {author} {\bibinfo {author} {\bibfnamefont {F.~G.}\ \bibnamefont
  {Tcheremissine}},\ }\bibfield  {title} {\enquote {\bibinfo {title} {{Direct
  numerical solution of the Boltzmann equation}},}\ }\href@noop {} {\bibfield
  {journal} {\bibinfo  {journal} {AIP Conf. Proc.}\ }\textbf {\bibinfo {volume}
  {762}},\ \bibinfo {pages} {677--685} (\bibinfo {year} {2005})}\BibitemShut
  {NoStop}%
\bibitem [{\citenamefont {Homolle}\ and\ \citenamefont
  {Hadjiconstantinou}({2007})}]{Homolle2007}%
  \BibitemOpen
  \bibfield  {author} {\bibinfo {author} {\bibfnamefont {T.~M.~M.}\
  \bibnamefont {Homolle}}\ and\ \bibinfo {author} {\bibfnamefont {N.~G.}\
  \bibnamefont {Hadjiconstantinou}},\ }\bibfield  {title} {\enquote {\bibinfo
  {title} {{A low-variance deviational simulation Monte Carlo for the Boltzmann
  equation}},}\ }\href {\doibase {10.1016/j.jcp.2007.07.006}} {\bibfield
  {journal} {\bibinfo  {journal} {J. Comput. Phys.}\ }\textbf {\bibinfo
  {volume} {{226}}},\ \bibinfo {pages} {{2341--2358}} (\bibinfo {year}
  {{2007}})}\BibitemShut {NoStop}%
\bibitem [{\citenamefont {Wu}\ \emph {et~al.}(2014{\natexlab{b}})\citenamefont
  {Wu}, \citenamefont {Reese},\ and\ \citenamefont {Zhang}}]{lei_Jfm}%
  \BibitemOpen
  \bibfield  {author} {\bibinfo {author} {\bibfnamefont {L.}~\bibnamefont
  {Wu}}, \bibinfo {author} {\bibfnamefont {J.~M.}\ \bibnamefont {Reese}}, \
  and\ \bibinfo {author} {\bibfnamefont {Y.~H.}\ \bibnamefont {Zhang}},\
  }\bibfield  {title} {\enquote {\bibinfo {title} {{Solving the Boltzmann
  equation by the fast spectral method: application to microflows}},}\
  }\href@noop {} {\bibfield  {journal} {\bibinfo  {journal} {J. Fluid Mech.}\
  }\textbf {\bibinfo {volume} {746}},\ \bibinfo {pages} {53--84} (\bibinfo
  {year} {2014}{\natexlab{b}})}\BibitemShut {NoStop}%
\bibitem [{\citenamefont {Bhatnagar}\ \emph {et~al.}(1954)\citenamefont
  {Bhatnagar}, \citenamefont {Gross},\ and\ \citenamefont
  {Krook}}]{Bhatnagar1954}%
  \BibitemOpen
  \bibfield  {author} {\bibinfo {author} {\bibfnamefont {P.~L.}\ \bibnamefont
  {Bhatnagar}}, \bibinfo {author} {\bibfnamefont {E.~P.}\ \bibnamefont
  {Gross}}, \ and\ \bibinfo {author} {\bibfnamefont {M.}~\bibnamefont
  {Krook}},\ }\bibfield  {title} {\enquote {\bibinfo {title} {A model for
  collision processes in gases. {I}. {S}mall amplitude processes in charged and
  neutral one-component systems},}\ }\href {\doibase 10.1103/PhysRev.94.511}
  {\bibfield  {journal} {\bibinfo  {journal} {Phys. Rev.}\ }\textbf {\bibinfo
  {volume} {94}},\ \bibinfo {pages} {511--525} (\bibinfo {year}
  {1954})}\BibitemShut {NoStop}%
\bibitem [{\citenamefont {Chu}(1965)}]{ChuDVM1965}%
  \BibitemOpen
  \bibfield  {author} {\bibinfo {author} {\bibfnamefont {C.~K.}\ \bibnamefont
  {Chu}},\ }\bibfield  {title} {\enquote {\bibinfo {title} {Kinetic-theoretic
  description of the formation of a shock wave},}\ }\href@noop {} {\bibfield
  {journal} {\bibinfo  {journal} {Phys. Fluids}\ }\textbf {\bibinfo {volume}
  {8}},\ \bibinfo {pages} {12} (\bibinfo {year} {1965})}\BibitemShut {NoStop}%
\bibitem [{\citenamefont {Yang}\ and\ \citenamefont
  {Huang}(1995)}]{YangJCP1995}%
  \BibitemOpen
  \bibfield  {author} {\bibinfo {author} {\bibfnamefont {J.~Y.}\ \bibnamefont
  {Yang}}\ and\ \bibinfo {author} {\bibfnamefont {J.~C.}\ \bibnamefont
  {Huang}},\ }\bibfield  {title} {\enquote {\bibinfo {title} {Rarefied flow
  computations using nonlinear model {Boltzmann} equations},}\ }\href@noop {}
  {\bibfield  {journal} {\bibinfo  {journal} {J. Comput. Phys.}\ }\textbf
  {\bibinfo {volume} {120}},\ \bibinfo {pages} {323--339} (\bibinfo {year}
  {1995})}\BibitemShut {NoStop}%
\bibitem [{\citenamefont {Sharipov}(1996)}]{Sharipov1996PF}%
  \BibitemOpen
  \bibfield  {author} {\bibinfo {author} {\bibfnamefont {F.}~\bibnamefont
  {Sharipov}},\ }\bibfield  {title} {\enquote {\bibinfo {title} {Rarefied gas
  flow through a slit. influence of the boundary condition},}\ }\href@noop {}
  {\bibfield  {journal} {\bibinfo  {journal} {Phys. Fluids}\ }\textbf {\bibinfo
  {volume} {8}},\ \bibinfo {pages} {262} (\bibinfo {year} {1996})}\BibitemShut
  {NoStop}%
\bibitem [{\citenamefont {Xu}\ and\ \citenamefont {Huang}(2010)}]{XuUGKS2010}%
  \BibitemOpen
  \bibfield  {author} {\bibinfo {author} {\bibfnamefont {K.}~\bibnamefont
  {Xu}}\ and\ \bibinfo {author} {\bibfnamefont {J.~C.}\ \bibnamefont {Huang}},\
  }\bibfield  {title} {\enquote {\bibinfo {title} {A unified gas-kinetic scheme
  for continuum and rarefied flows},}\ }\href@noop {} {\bibfield  {journal}
  {\bibinfo  {journal} {J. Comput. Phys.}\ }\textbf {\bibinfo {volume} {229}},\
  \bibinfo {pages} {7747--7764} (\bibinfo {year} {2010})}\BibitemShut {NoStop}%
\bibitem [{\citenamefont {Guo}\ \emph {et~al.}(2013)\citenamefont {Guo},
  \citenamefont {Xu},\ and\ \citenamefont {Wang}}]{GuoDUGKS2013}%
  \BibitemOpen
  \bibfield  {author} {\bibinfo {author} {\bibfnamefont {Z.}~\bibnamefont
  {Guo}}, \bibinfo {author} {\bibfnamefont {K.}~\bibnamefont {Xu}}, \ and\
  \bibinfo {author} {\bibfnamefont {R.~J.}\ \bibnamefont {Wang}},\ }\bibfield
  {title} {\enquote {\bibinfo {title} {Discrete unified gas kinetic scheme for
  all {Knudsen} number flows: {Low-speed} isothermal case},}\ }\href@noop {}
  {\bibfield  {journal} {\bibinfo  {journal} {Phys. Rev. E}\ }\textbf {\bibinfo
  {volume} {88}},\ \bibinfo {pages} {033305} (\bibinfo {year}
  {2013})}\BibitemShut {NoStop}%
\bibitem [{\citenamefont {S.Taguchi}\ and\ \citenamefont
  {Aoki}(2012)}]{AokiJFM2012}%
  \BibitemOpen
  \bibfield  {author} {\bibinfo {author} {\bibnamefont {S.Taguchi}}\ and\
  \bibinfo {author} {\bibfnamefont {K.}~\bibnamefont {Aoki}},\ }\bibfield
  {title} {\enquote {\bibinfo {title} {Rarefied gas flow around a sharp edge
  induced by a temperature field},}\ }\href@noop {} {\bibfield  {journal}
  {\bibinfo  {journal} {J. Fluid Mech.}\ }\textbf {\bibinfo {volume} {694}},\
  \bibinfo {pages} {191--224} (\bibinfo {year} {2012})}\BibitemShut {NoStop}%
\bibitem [{\citenamefont {He}\ and\ \citenamefont {Luo}(1997)}]{Xiaoyi1997}%
  \BibitemOpen
  \bibfield  {author} {\bibinfo {author} {\bibfnamefont {X.}~\bibnamefont
  {He}}\ and\ \bibinfo {author} {\bibfnamefont {L.~S.}\ \bibnamefont {Luo}},\
  }\bibfield  {title} {\enquote {\bibinfo {title} {{Theory of the lattice
  Boltzmann method: From the Boltzmann equation to the lattice Boltzmann
  equation}},}\ }\href@noop {} {\bibfield  {journal} {\bibinfo  {journal}
  {Phys. Rev. E}\ }\textbf {\bibinfo {volume} {56}},\ \bibinfo {pages}
  {6811--6817} (\bibinfo {year} {1997})}\BibitemShut {NoStop}%
\bibitem [{\citenamefont {Shan}\ \emph {et~al.}(2006)\citenamefont {Shan},
  \citenamefont {Yuan},\ and\ \citenamefont {Chen}}]{ShanJFM2006}%
  \BibitemOpen
  \bibfield  {author} {\bibinfo {author} {\bibfnamefont {X.}~\bibnamefont
  {Shan}}, \bibinfo {author} {\bibfnamefont {X.}~\bibnamefont {Yuan}}, \ and\
  \bibinfo {author} {\bibfnamefont {H.}~\bibnamefont {Chen}},\ }\bibfield
  {title} {\enquote {\bibinfo {title} {Kinetic theory representation of
  hydrodynamics: a way beyond the {Navier-Stokes} equation},}\ }\href@noop {}
  {\bibfield  {journal} {\bibinfo  {journal} {J. Fluid Mech.}\ }\textbf
  {\bibinfo {volume} {550}},\ \bibinfo {pages} {413--441} (\bibinfo {year}
  {2006})}\BibitemShut {NoStop}%
\bibitem [{\citenamefont {Kim}\ \emph {et~al.}(2008)\citenamefont {Kim},
  \citenamefont {Pitsch},\ and\ \citenamefont {Boyd}}]{KimJcp2008}%
  \BibitemOpen
  \bibfield  {author} {\bibinfo {author} {\bibfnamefont {S.~H.}\ \bibnamefont
  {Kim}}, \bibinfo {author} {\bibfnamefont {H.}~\bibnamefont {Pitsch}}, \ and\
  \bibinfo {author} {\bibfnamefont {I.~D.}\ \bibnamefont {Boyd}},\ }\bibfield
  {title} {\enquote {\bibinfo {title} {{Accuracy of higher-order lattice
  Boltzmann methods for microscale flows with finite Knudsen numbers}},}\
  }\href@noop {} {\bibfield  {journal} {\bibinfo  {journal} {J. Comput. Phys.}\
  }\textbf {\bibinfo {volume} {227}},\ \bibinfo {pages} {8655--8671} (\bibinfo
  {year} {2008})}\BibitemShut {NoStop}%
\bibitem [{\citenamefont {Meng}\ and\ \citenamefont
  {Zhang}(2011{\natexlab{a}})}]{MengJcp2011}%
  \BibitemOpen
  \bibfield  {author} {\bibinfo {author} {\bibfnamefont {J.}~\bibnamefont
  {Meng}}\ and\ \bibinfo {author} {\bibfnamefont {Y.~H.}\ \bibnamefont
  {Zhang}},\ }\bibfield  {title} {\enquote {\bibinfo {title} {{Accuracy
  analysis of high-order lattice Boltzmann models for rarefied gas flows}},}\
  }\href@noop {} {\bibfield  {journal} {\bibinfo  {journal} {J. Comput. Phys.}\
  }\textbf {\bibinfo {volume} {230}},\ \bibinfo {pages} {835--849} (\bibinfo
  {year} {2011}{\natexlab{a}})}\BibitemShut {NoStop}%
\bibitem [{\citenamefont {Meng}\ and\ \citenamefont
  {Zhang}(2011{\natexlab{b}})}]{MengPre2011}%
  \BibitemOpen
  \bibfield  {author} {\bibinfo {author} {\bibfnamefont {J.}~\bibnamefont
  {Meng}}\ and\ \bibinfo {author} {\bibfnamefont {Y.~H.}\ \bibnamefont
  {Zhang}},\ }\bibfield  {title} {\enquote {\bibinfo {title} {{Gauss-Hermite
  quadratures and accuracy of lattice Boltzmann models for non-equilibrium gas
  flows}},}\ }\href@noop {} {\bibfield  {journal} {\bibinfo  {journal} {Phys.
  Rev. E}\ }\textbf {\bibinfo {volume} {83}},\ \bibinfo {pages} {036704}
  (\bibinfo {year} {2011}{\natexlab{b}})}\BibitemShut {NoStop}%
\bibitem [{\citenamefont {Shi}\ \emph {et~al.}(2011)\citenamefont {Shi},
  \citenamefont {Brookes}, \citenamefont {Yap},\ and\ \citenamefont
  {Sader}}]{ShiPreR2011}%
  \BibitemOpen
  \bibfield  {author} {\bibinfo {author} {\bibfnamefont {Y.}~\bibnamefont
  {Shi}}, \bibinfo {author} {\bibfnamefont {P.~L.}\ \bibnamefont {Brookes}},
  \bibinfo {author} {\bibfnamefont {Y.~W.}\ \bibnamefont {Yap}}, \ and\
  \bibinfo {author} {\bibfnamefont {J.~E.}\ \bibnamefont {Sader}},\ }\bibfield
  {title} {\enquote {\bibinfo {title} {{Accuracy of the lattice Boltzmann
  method for low-speed noncontinuum flows}},}\ }\href@noop {} {\bibfield
  {journal} {\bibinfo  {journal} {Phys. Rev. E}\ }\textbf {\bibinfo {volume}
  {83}},\ \bibinfo {pages} {045701(R)} (\bibinfo {year} {2011})}\BibitemShut
  {NoStop}%
\bibitem [{\citenamefont {Ambru\c{s}}\ and\ \citenamefont
  {Sofonea}(2012)}]{SofoneaPre2012}%
  \BibitemOpen
  \bibfield  {author} {\bibinfo {author} {\bibfnamefont {V.~E.}\ \bibnamefont
  {Ambru\c{s}}}\ and\ \bibinfo {author} {\bibfnamefont {V.}~\bibnamefont
  {Sofonea}},\ }\bibfield  {title} {\enquote {\bibinfo {title} {{High-order
  thermal lattice Boltzmann models derived by means of Gauss quadrature in the
  spherical coordinate system}},}\ }\href@noop {} {\bibfield  {journal}
  {\bibinfo  {journal} {Phys. Rev. E}\ }\textbf {\bibinfo {volume} {86}},\
  \bibinfo {pages} {016708} (\bibinfo {year} {2012})}\BibitemShut {NoStop}%
\bibitem [{\citenamefont {Ambru\c{s}}\ and\ \citenamefont
  {Sofonea}(2014)}]{SofoneaPre2014}%
  \BibitemOpen
  \bibfield  {author} {\bibinfo {author} {\bibfnamefont {V.~E.}\ \bibnamefont
  {Ambru\c{s}}}\ and\ \bibinfo {author} {\bibfnamefont {V.}~\bibnamefont
  {Sofonea}},\ }\bibfield  {title} {\enquote {\bibinfo {title} {{Implementation
  of diffuse-reflection boundary conditions using lattice Boltzmann models
  based on half-space Gauss-Laguerre quadratures}},}\ }\href@noop {} {\bibfield
   {journal} {\bibinfo  {journal} {Phys. Rev. E}\ }\textbf {\bibinfo {volume}
  {89}},\ \bibinfo {pages} {041301} (\bibinfo {year} {2014})}\BibitemShut
  {NoStop}%
\bibitem [{\citenamefont {Shi}\ \emph {et~al.}(2015)\citenamefont {Shi},
  \citenamefont {Yap},\ and\ \citenamefont {Sader}}]{ShiPre2015}%
  \BibitemOpen
  \bibfield  {author} {\bibinfo {author} {\bibfnamefont {Y.}~\bibnamefont
  {Shi}}, \bibinfo {author} {\bibfnamefont {Y.~W.}\ \bibnamefont {Yap}}, \ and\
  \bibinfo {author} {\bibfnamefont {J.~E.}\ \bibnamefont {Sader}},\ }\bibfield
  {title} {\enquote {\bibinfo {title} {{Linearized lattice Boltzmann method for
  micro- and nanoscale flow and heat transfer}},}\ }\href@noop {} {\bibfield
  {journal} {\bibinfo  {journal} {Phys. Rev. E}\ }\textbf {\bibinfo {volume}
  {92}},\ \bibinfo {pages} {013307} (\bibinfo {year} {2015})}\BibitemShut
  {NoStop}%
\bibitem [{\citenamefont {Zhang}\ \emph {et~al.}(2006)\citenamefont {Zhang},
  \citenamefont {Gu}, \citenamefont {Barber},\ and\ \citenamefont
  {Emerson}}]{ZhangPre2006}%
  \BibitemOpen
  \bibfield  {author} {\bibinfo {author} {\bibfnamefont {Y.~H.}\ \bibnamefont
  {Zhang}}, \bibinfo {author} {\bibfnamefont {X.~J.}\ \bibnamefont {Gu}},
  \bibinfo {author} {\bibfnamefont {R.~W.}\ \bibnamefont {Barber}}, \ and\
  \bibinfo {author} {\bibfnamefont {D.~R.}\ \bibnamefont {Emerson}},\
  }\bibfield  {title} {\enquote {\bibinfo {title} {Capturing {Knudsen} layer
  phenomena using a lattice {Boltzmann} model},}\ }\href@noop {} {\bibfield
  {journal} {\bibinfo  {journal} {Phys. Rev. E}\ }\textbf {\bibinfo {volume}
  {74}},\ \bibinfo {pages} {046704} (\bibinfo {year} {2006})}\BibitemShut
  {NoStop}%
\bibitem [{\citenamefont {Zhang}\ \emph {et~al.}(2007)\citenamefont {Zhang},
  \citenamefont {Gu}, \citenamefont {Barber},\ and\ \citenamefont
  {Emerson}}]{ZhangEpl2007}%
  \BibitemOpen
  \bibfield  {author} {\bibinfo {author} {\bibfnamefont {Y.~H.}\ \bibnamefont
  {Zhang}}, \bibinfo {author} {\bibfnamefont {X.~J.}\ \bibnamefont {Gu}},
  \bibinfo {author} {\bibfnamefont {R.~W.}\ \bibnamefont {Barber}}, \ and\
  \bibinfo {author} {\bibfnamefont {D.~R.}\ \bibnamefont {Emerson}},\
  }\bibfield  {title} {\enquote {\bibinfo {title} {{Modelling thermal flow in
  the transition regime using a lattice Boltzmann approach}},}\ }\href@noop {}
  {\bibfield  {journal} {\bibinfo  {journal} {Europhys. Lett.}\ }\textbf
  {\bibinfo {volume} {77}},\ \bibinfo {pages} {30003} (\bibinfo {year}
  {2007})}\BibitemShut {NoStop}%
\bibitem [{\citenamefont {Tang}\ \emph {et~al.}(2008)\citenamefont {Tang},
  \citenamefont {Gu}, \citenamefont {Barber}, \citenamefont {Emerson},\ and\
  \citenamefont {Zhang}}]{Tang2008}%
  \BibitemOpen
  \bibfield  {author} {\bibinfo {author} {\bibfnamefont {G.~H.}\ \bibnamefont
  {Tang}}, \bibinfo {author} {\bibfnamefont {X.~J.}\ \bibnamefont {Gu}},
  \bibinfo {author} {\bibfnamefont {R.~W.}\ \bibnamefont {Barber}}, \bibinfo
  {author} {\bibfnamefont {D.~R.}\ \bibnamefont {Emerson}}, \ and\ \bibinfo
  {author} {\bibfnamefont {Y.~H.}\ \bibnamefont {Zhang}},\ }\bibfield  {title}
  {\enquote {\bibinfo {title} {{Lattice Boltzmann simulation of nonequilibrium
  effects in oscillatory gas flow}},}\ }\href@noop {} {\bibfield  {journal}
  {\bibinfo  {journal} {Phys. Rev. E}\ }\textbf {\bibinfo {volume} {78}},\
  \bibinfo {pages} {026706} (\bibinfo {year} {2008})}\BibitemShut {NoStop}%
\bibitem [{\citenamefont {Guo}\ \emph {et~al.}(2008)\citenamefont {Guo},
  \citenamefont {Zheng},\ and\ \citenamefont {Shi}}]{Guo2008Slip}%
  \BibitemOpen
  \bibfield  {author} {\bibinfo {author} {\bibfnamefont {Z.}~\bibnamefont
  {Guo}}, \bibinfo {author} {\bibfnamefont {C.}~\bibnamefont {Zheng}}, \ and\
  \bibinfo {author} {\bibfnamefont {B.}~\bibnamefont {Shi}},\ }\bibfield
  {title} {\enquote {\bibinfo {title} {{Lattice Boltzmann equation with
  multiple effective relaxation times for gaseous microscale flow}},}\
  }\href@noop {} {\bibfield  {journal} {\bibinfo  {journal} {Phys. Rev. E}\
  }\textbf {\bibinfo {volume} {77}},\ \bibinfo {pages} {036707} (\bibinfo
  {year} {2008})}\BibitemShut {NoStop}%
\bibitem [{\citenamefont {Li}\ \emph {et~al.}(2011)\citenamefont {Li},
  \citenamefont {He}, \citenamefont {Tang},\ and\ \citenamefont
  {Tao}}]{QliLBM2011}%
  \BibitemOpen
  \bibfield  {author} {\bibinfo {author} {\bibfnamefont {Q.}~\bibnamefont
  {Li}}, \bibinfo {author} {\bibfnamefont {Y.~L.}\ \bibnamefont {He}}, \bibinfo
  {author} {\bibfnamefont {G.~H.}\ \bibnamefont {Tang}}, \ and\ \bibinfo
  {author} {\bibfnamefont {W.~Q.}\ \bibnamefont {Tao}},\ }\bibfield  {title}
  {\enquote {\bibinfo {title} {{Lattice Boltzmann modeling of microchannel
  flows in the transition flow regime}},}\ }\href@noop {} {\bibfield  {journal}
  {\bibinfo  {journal} {{Microfluidics Nanofluidics}}\ }\textbf {\bibinfo
  {volume} {10}},\ \bibinfo {pages} {607--618} (\bibinfo {year}
  {2011})}\BibitemShut {NoStop}%
\bibitem [{\citenamefont {Yuan}\ and\ \citenamefont
  {Rahman}(2016)}]{Rahman2016}%
  \BibitemOpen
  \bibfield  {author} {\bibinfo {author} {\bibfnamefont {Y.}~\bibnamefont
  {Yuan}}\ and\ \bibinfo {author} {\bibfnamefont {S.}~\bibnamefont {Rahman}},\
  }\bibfield  {title} {\enquote {\bibinfo {title} {{Extended application of
  lattice Boltzmann method to rarefied gas flow in micro-channels}},}\
  }\href@noop {} {\bibfield  {journal} {\bibinfo  {journal} {Physica A}\
  }\textbf {\bibinfo {volume} {463}},\ \bibinfo {pages} {25--36} (\bibinfo
  {year} {2016})}\BibitemShut {NoStop}%
\bibitem [{\citenamefont {Deng}\ \emph {et~al.}(2016)\citenamefont {Deng},
  \citenamefont {Chen},\ and\ \citenamefont {Shao}}]{ChenRough2016}%
  \BibitemOpen
  \bibfield  {author} {\bibinfo {author} {\bibfnamefont {Z.}~\bibnamefont
  {Deng}}, \bibinfo {author} {\bibfnamefont {Y.}~\bibnamefont {Chen}}, \ and\
  \bibinfo {author} {\bibfnamefont {C.}~\bibnamefont {Shao}},\ }\bibfield
  {title} {\enquote {\bibinfo {title} {Gas flow through rough microchannels in
  the transition flow regime},}\ }\href@noop {} {\bibfield  {journal} {\bibinfo
   {journal} {Phys. Rev. E}\ }\textbf {\bibinfo {volume} {93}},\ \bibinfo
  {pages} {013128} (\bibinfo {year} {2016})}\BibitemShut {NoStop}%
\bibitem [{\citenamefont {Zhao}\ \emph {et~al.}(2016)\citenamefont {Zhao},
  \citenamefont {Yao}, \citenamefont {Li}, \citenamefont {Zhang}, \citenamefont
  {Zhang}, \citenamefont {Yang},\ and\ \citenamefont {Sun}}]{YaoJAP2016}%
  \BibitemOpen
  \bibfield  {author} {\bibinfo {author} {\bibfnamefont {J.}~\bibnamefont
  {Zhao}}, \bibinfo {author} {\bibfnamefont {J.}~\bibnamefont {Yao}}, \bibinfo
  {author} {\bibfnamefont {A.}~\bibnamefont {Li}}, \bibinfo {author}
  {\bibfnamefont {M.}~\bibnamefont {Zhang}}, \bibinfo {author} {\bibfnamefont
  {L.}~\bibnamefont {Zhang}}, \bibinfo {author} {\bibfnamefont
  {Y.}~\bibnamefont {Yang}}, \ and\ \bibinfo {author} {\bibfnamefont
  {H.}~\bibnamefont {Sun}},\ }\bibfield  {title} {\enquote {\bibinfo {title}
  {Simulation of microscale gas flow in heterogeneous porous media based on the
  lattice {Boltzmann} method},}\ }\href@noop {} {\bibfield  {journal} {\bibinfo
   {journal} {J. Appl. Phys.}\ }\textbf {\bibinfo {volume} {120}},\ \bibinfo
  {pages} {084306} (\bibinfo {year} {2016})}\BibitemShut {NoStop}%
\bibitem [{\citenamefont {Guo}\ \emph {et~al.}(2014)\citenamefont {Guo},
  \citenamefont {Qin},\ and\ \citenamefont {Zheng}}]{GuoPRE2014}%
  \BibitemOpen
  \bibfield  {author} {\bibinfo {author} {\bibfnamefont {Z.}~\bibnamefont
  {Guo}}, \bibinfo {author} {\bibfnamefont {J.}~\bibnamefont {Qin}}, \ and\
  \bibinfo {author} {\bibfnamefont {C.}~\bibnamefont {Zheng}},\ }\bibfield
  {title} {\enquote {\bibinfo {title} {Generalized second-order slip boundary
  condition for nonequilibrium gas flows},}\ }\href@noop {} {\bibfield
  {journal} {\bibinfo  {journal} {Phys. Rev. E}\ }\textbf {\bibinfo {volume}
  {89}},\ \bibinfo {pages} {013021} (\bibinfo {year} {2014})}\BibitemShut
  {NoStop}%
\bibitem [{\citenamefont {Sharipov}\ and\ \citenamefont
  {Bertoldo}(2009)}]{Sharipov2009}%
  \BibitemOpen
  \bibfield  {author} {\bibinfo {author} {\bibfnamefont {F.}~\bibnamefont
  {Sharipov}}\ and\ \bibinfo {author} {\bibfnamefont {G.}~\bibnamefont
  {Bertoldo}},\ }\bibfield  {title} {\enquote {\bibinfo {title} {{Poiseuille
  flow and thermal creep based on the Boltzmann equation with the Lennard-Jones
  potential over a wide range of the Knudsen number}},}\ }\href@noop {}
  {\bibfield  {journal} {\bibinfo  {journal} {Phys. Fluids}\ }\textbf {\bibinfo
  {volume} {21}},\ \bibinfo {pages} {067101} (\bibinfo {year}
  {2009})}\BibitemShut {NoStop}%
\bibitem [{\citenamefont {Doi}(2010)}]{Doi2010}%
  \BibitemOpen
  \bibfield  {author} {\bibinfo {author} {\bibfnamefont {T.}~\bibnamefont
  {Doi}},\ }\bibfield  {title} {\enquote {\bibinfo {title} {{Numerical analysis
  of the Poiseuille flow and thermal transpiration of a rarefied gas through a
  pipe with a rectangular cross section based on the linearized Boltzmann
  equation for a hard sphere molecular gas}},}\ }\href@noop {} {\bibfield
  {journal} {\bibinfo  {journal} {J. Vac. Sci. Technol. A}\ }\textbf {\bibinfo
  {volume} {28}},\ \bibinfo {pages} {603--612} (\bibinfo {year}
  {2010})}\BibitemShut {NoStop}%
\bibitem [{\citenamefont {Sharipov}\ and\ \citenamefont
  {Graur}(2012)}]{GraurVacuum2012}%
  \BibitemOpen
  \bibfield  {author} {\bibinfo {author} {\bibfnamefont {F.}~\bibnamefont
  {Sharipov}}\ and\ \bibinfo {author} {\bibfnamefont {I.~A.}\ \bibnamefont
  {Graur}},\ }\bibfield  {title} {\enquote {\bibinfo {title} {Rarefied gas flow
  through a zigzag channel},}\ }\href@noop {} {\bibfield  {journal} {\bibinfo
  {journal} {Vacuum}\ }\textbf {\bibinfo {volume} {86}},\ \bibinfo {pages}
  {1778--1782} (\bibinfo {year} {2012})}\BibitemShut {NoStop}%
\bibitem [{\citenamefont {Maxwell}(1879)}]{Maxwell1879}%
  \BibitemOpen
  \bibfield  {author} {\bibinfo {author} {\bibfnamefont {J.~C.}\ \bibnamefont
  {Maxwell}},\ }\bibfield  {title} {\enquote {\bibinfo {title} {On stresses in
  rarefied gases arising from inequalities of temperature},}\ }\href@noop {}
  {\bibfield  {journal} {\bibinfo  {journal} {Philosophical Transactions of the
  Royal Society Part 1}\ }\textbf {\bibinfo {volume} {170}},\ \bibinfo {pages}
  {231--256} (\bibinfo {year} {1879})}\BibitemShut {NoStop}%
\bibitem [{\citenamefont {Wu}\ \emph {et~al.}(2017)\citenamefont {Wu},
  \citenamefont {Ho}, \citenamefont {Germanou}, \citenamefont {Gu},
  \citenamefont {Liu}, \citenamefont {Xu},\ and\ \citenamefont
  {Zhang}}]{LeiComment2017}%
  \BibitemOpen
  \bibfield  {author} {\bibinfo {author} {\bibfnamefont {L.}~\bibnamefont
  {Wu}}, \bibinfo {author} {\bibfnamefont {M.~T.}\ \bibnamefont {Ho}}, \bibinfo
  {author} {\bibfnamefont {L.}~\bibnamefont {Germanou}}, \bibinfo {author}
  {\bibfnamefont {X.~J.}\ \bibnamefont {Gu}}, \bibinfo {author} {\bibfnamefont
  {C.}~\bibnamefont {Liu}}, \bibinfo {author} {\bibfnamefont {K.}~\bibnamefont
  {Xu}}, \ and\ \bibinfo {author} {\bibfnamefont {Y.~H.}\ \bibnamefont
  {Zhang}},\ }\bibfield  {title} {\enquote {\bibinfo {title} {{A comment on `An
  improved macroscale model for gas slip flow in porous media'}},}\ }\href@noop
  {} {\bibfield  {journal} {\bibinfo  {journal} {Arxiv:1702.04102}\ } (\bibinfo
  {year} {2017})}\BibitemShut {NoStop}%
\end{thebibliography}%

\end{document}